\documentclass[%
 aip,
 jmp,%
 amsmath,amssymb,
 reprint,%
]{revtex4-1}

\usepackage[dvipdfmx]{graphicx}

\usepackage{dcolumn}
\usepackage{bm}
\usepackage{amssymb,amsfonts,amsmath}
\usepackage{color}
\usepackage{amsmath,amssymb,amscd,ulem}
\usepackage{comment}
\usepackage{framed}
\usepackage{setspace}

\newcommand{\ProbP}{\mathbb{P}}
\newcommand{\ProbQ}{\mathbb{Q}}

\newcommand{\ProbT}{\mathbb{T}}

\newcommand{\LF}{K}
\newcommand{\lf}{k}
\newcommand{\defeq}{:=}

\newcommand{\CMF}{\Psi}

\newcommand{\MI}{\mathcal{I}}
\newcommand{\average}[1]{\left< #1 \right>}

\newcommand{\KLD}{\mathcal{D}}
\newcommand{\KL}[2]{\KLD[#1\|#2]}

 \DeclareMathAlphabet{\mathpzc}{OT1}{pzc}{m}{it}\newcommand{\xpzc}{\mathpzc{X}}
\newcommand{\ypzc}{\mathpzc{Y}}
\newcommand{\zpzc}{\mathpzc{Z}}

\newcommand{\popN}{\mathcal{N}}
\newcommand{\com}[1]{#1}
\newcommand{\eqnref}[1]{eq. (\ref{#1})}

\newcommand{\setx}{\mathfrak{S}^{x}}
\newcommand{\sety}{\mathfrak{S}^{y}}
\newcommand{\setz}{\mathfrak{S}^{z}}
\newcommand{\statx}{\mathfrak{s}^{x}}
\newcommand{\staty}{\mathfrak{s}^{y}}
\newcommand{\statz}{\mathfrak{s}^{z}}

\newcommand{\loss}{\mathrm{loss}}

\newcommand{\ISgain}{\mathcal{G}}
\newcommand{\isgain}{g}

\newcommand{\Dt}{\Delta t}

\begin{document}

\preprint{AIP/123-QED}

\title[Fitness Gain of Individual Sensing]{Individual Sensing can Gain more Fitness than its Information}

\author{Tetsuya J. Kobayashi}
 \email{tetsuya@mail.crmind.net}
 \homepage{http://research.crmind.net}
 \affiliation{Institute of Industrial Science, The University of Tokyo, 4-6-1 Komaba, Meguro-ku 153-8505, Tokyo, Japan}
 \affiliation{PREST, Japan Science and Technology Agency (JST), 4-1-8 Honcho Kawaguchi, Saitama 332-0012, Japan}
\author{Yuki Sughiyama}%
 \affiliation{Institute of Industrial Science, The University of Tokyo, 4-6-1 Komaba, Meguro-ku 153-8505, Tokyo, Japan}

\date{\today}

\begin{abstract}
Mutual information and its causal variant, directed information, have been widely used to quantitatively characterize the performance of biological sensing and information transduction.
However, once coupled with selection in response to decision-making, the sensing signal could have more or less evolutionary value than its mutual or directed information.
In this work, we show that an individually sensed signal always has a better fitness value, on average, than its mutual or directed information. 
The fitness gain, which satisfies fluctuation relations (FRs), is attributed to the selection of organisms in a population that obtain a better sensing signal by chance.
A new quantity, similar to the coarse-grained entropy production in information thermodynamics, is introduced to quantify the total fitness gain from individual sensing, which also satisfies FRs.
Using this quantity, the optimizing fitness gain from individual sensing is shown to be related to fidelity allocations for individual environmental histories.
Our results are supplemented by numerical verifications of FRs, and a discussion on how this problem is linked to information encoding and decoding.
\end{abstract}

\pacs{Valid PACS appear here}
\maketitle


\section{Introduction}
Most biological systems are equipped with active sensing machinery to monitor the ever-changing environment. 
The fidelity of sensing is crucial to choosing appropriate states and behaviors in response to changes in environmental states\cite{Perkins:2009cg,Kobayashi:2012ji,Bowsher:2014cdb}.
Instantaneous mutual information, path-wise mutual information, and its causal variant, directed information, have been used to quantitatively characterize the performance of the sensing and information transduction, theoretically\cite{Tostevin:2009ja,Kobayashi:2010vo,Bowsher:2012kb} and experimentally\cite{Tkacik:2008dqa,Cheong:2011jpa,Brennan:2012cj,Uda:2013cfa}.
These information measures are also fundamental to the thermodynamic cost of sensing\cite{Barato:2014df,Das:2016vt}.

However, it is still elusive whether these measures can appropriately quantify the biological and fitness value of sensed information.
Despite intensive works on the fitness value of information\cite{Haccou:1995tf,Bergstrom:2004um,Kussell:2005dg,DonaldsonMatasci:2010ie,Rivoire:2011fy,Pugatch:2013va,Rivoire:2014kt,Rivoire:2015if}, almost all works considered a biologically unrealistic situation in which all cells or organisms in a population receive a common sensing signal, which is the requisite for proving that the fitness value of sensing is bounded by the information measures.
Few studies have conjectured that biologically realistic sensing by individual organisms may have greater fitness value than these measures\cite{Rivoire:2011fy,Rivoire:2015if}.

In this work, we resolve this problem by generally proving that the individual sensing always has greater fitness value than common sensing does. 
The additional fitness gain, which satisfies fluctuation relations (FRs), is attributed to the selection of organisms that obtains a correct sensing signal by chance.
A new quantity, which is similar to the coarse-grained entropy production in information thermodynamics, is introduced to quantify the total fitness gain from the individual sensing, the upper bound of which is strictly higher than the directed information.
We further show that the optimization of this quantity is closely related to optimizing an auto-encoding network, in which sensing, phenotypic switching, and metabolic allocation work as encoding, processing, and decoding, respectively.
Our general results, especially those for FRs, are verified by a numerical simulation.


\begin{figure}
\includegraphics[width=\linewidth]{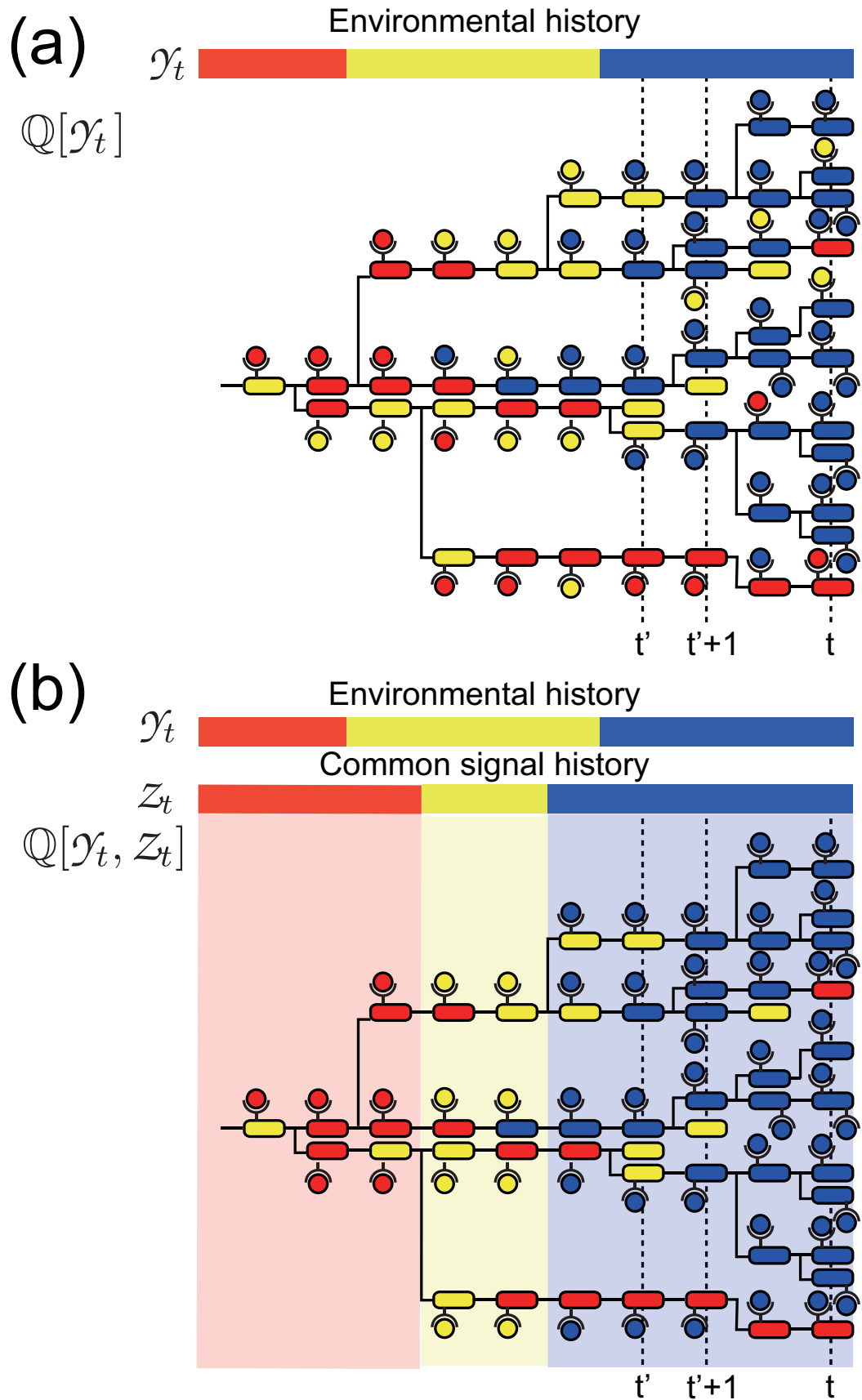}
\caption{\label{fig1} Schematic diagrams of population dynamics of cells with individual (a) and common (b) sensing.
The colors of cells and molecules on the cells represent phenotypic states and sensing signal, respectively. Bars on the diagrams indicate the histories of environmental states and common sensing.}
\end{figure}

\section{Modeling sensing and adaptation processes}
We consider a population of an asexual organism that replicates with an instantaneous replication rate $\lf(x,y)$, depending on its phenotype $x\in \setx$ and the state of environment $y\in \sety$, where the phenotypic and environmental states are assumed to be discrete and finite, for simplicity.
The organism switches its phenotype stochastically from $x$ to $x'$ by exploiting sensing signal $z\in \setz$ with a transition probability $\ProbT_{F}(x'|x,z)$ within a small time interval $\Dt$.
Depending on the physical entity of $z$, the sensing can be categorized as either individual or common sensing\cite{Rivoire:2011fy,Rivoire:2015if}.
In the case of individual sensing, $z$ is the state of a sensing system of the organism, such as the activity of receptors.
Because of stochasticity in the sensing process, the individual organisms receive different sensing signals $z$ (Fig.1 (a)). 
By assuming that the stochastic sensing output $z$ depends on the state of the environment $y$ as $\ProbT_{S}(z|y)$, we describe the dynamics of the number of organisms $\popN_{t}^{\ypzc}(x_{t},z_{t})$ that have phenotypic state $x_{t}$ with sensing signal $z_{t}$ at $t$ as 
\begin{align}
&\popN_{t+1}^{\ypzc}(x_{t+1},z_{t+1})=e^{\lf(x_{t+1},y_{t+1})}\label{eq:eqInd} \\
& \times \sum_{x_{t},z_{t}}\ProbT_{F}(x_{t+1}|x_{t},z_{t+1})\ProbT_{S}(z_{t+1}|z_{t},y_{t+1})\popN_{t}^{\ypzc}(x_{t},z_{t}),\notag
\end{align}
where $\ypzc_{t} \defeq \{y_{0},\cdots, y_{t}\}$ is the history of the environmental state, the statistical properties of which are characterized by path probability $\ProbQ[\ypzc_{t}]$.

In contrast, in the case of common sensing, $z$ is assumed to be partial information on the environmental state that is common to all organisms\cite{Kobayashi:2015eca,Kobayashi:2017tk} (Fig.1 (b)). 
An example is an extracellularl chemical that correlates with the environmental state and can be sensed by the organisms with negligible error.  The dynamics of the number of organisms $\popN_{t}^{\ypzc,\zpzc}(x_{t})$ with phenotypic state $x$ at time $t$ under a realization of environmental and common signal histories, $\ypzc_{t}$ and $\zpzc_{t}$, can be represented as
\begin{align}
\popN_{t+1}^{\ypzc,\zpzc}(x_{t+1}) =&e^{\lf(x_{t+1},y_{t+1})} \label{eq:eqCom}\\
& \times \sum_{x_{t}\in \setx}\ProbT_{F}(x_{t+1}|x_{t},z_{t+1})\popN_{t}^{\ypzc,\zpzc}(x_{t}). \notag
\end{align}
We assume that the history of the common signal $\zpzc_{t} \defeq \{z_{0},\cdots, z_{t}\}$ follows a statistical law $\ProbQ[\zpzc_{t}\|\ypzc_{t}]$, which is causally conditional on the environmental history.
While common sensing is not biologically realistic enough, most previous works on the fitness value of information only addressed common sensing, and prove that the fitness gain of common sensing is upper bounded by the directed information \cite{Kobayashi:2015eca,Kobayashi:2017tk}.

\subsection{Fitness of a population with individual and common sensing}
The fitness of a population with individual sensing $\CMF^{i}[\ypzc_{t}]$ and with common sensing $\CMF^{c}[\ypzc_{t},\zpzc_{t}]$ can be defined respectively as 
\begin{align}
\CMF^{i}[\ypzc_{t}] \defeq \ln\frac{\popN_{t}^{\ypzc}}{\popN_{0}^{\ypzc}},\quad \CMF^{c}[\ypzc_{t},\zpzc_{t}] \defeq \ln\frac{\popN_{t}^{\ypzc,\zpzc}}{\popN_{0}^{\ypzc,\zpzc}},
\end{align}
where $\popN_{t}^{\ypzc}:=\sum_{x_{t},z_{t}}\popN_{t}^{\ypzc}(x_{t}, z_{t})$ and $\popN_{t}^{\ypzc, \zpzc}:=\sum_{x_{t}}\popN_{t}^{\ypzc, \zpzc}(x_{t})$.
By defining a pathwise historical fitness\cite{Leibler:2010jx} 
\begin{align}
\LF[\xpzc_{t},\ypzc_{t}]\defeq \sum_{\tau=0}^{t-1}\lf(x_{\tau+1}, y_{\tau+1}),
\end{align}
 and path probabilities for phenotypic and signal histories 
 \begin{align}
 \ProbP_{F}[\xpzc_{t}\|\zpzc_{t}] &\defeq \left[\prod_{\tau=0}^{t-1}\ProbT_{F}(x_{\tau+1}|x_{\tau},z_{\tau+1})\right]p_{F}(x_{0}),\\
 \ProbP_{S}[\zpzc_{t}\|\ypzc_{t}] &\defeq \left[\prod_{\tau=0}^{t-1}\ProbT_{S}(z_{\tau+1}|z_{\tau},y_{\tau+1})\right]p_{S}(z_{0}|y_{0}),
 \end{align}
respectively.
In conjunction with eqns (\ref{eq:eqInd}) and (\ref{eq:eqCom}),  we can explicitly represent the fitnesses\cite{Leibler:2010jx,Sughiyama:2015cf,Kobayashi:2015eca,Kobayashi:2017tk} as
\begin{align*}
\CMF^{i}[\ypzc_{t}] &= \ln\average{e^{\LF[\xpzc_{t},\ypzc_{t}]}}_{\ProbP_{F,S}[\xpzc_{t}|\ypzc_{t}]},\\
\CMF^{c}[\ypzc_{t},\zpzc_{t}] &=\ln\average{e^{\LF[\xpzc_{t},\ypzc_{t}]}}_{\ProbP_{F}[\xpzc_{t}\|\zpzc_{t}]},
\end{align*}
 where $\average{\cdot}_{\ProbP[\xpzc_{t}]}$ is the average with respect to $\ProbP[\xpzc_{t}]$, and $\ProbP_{F,S}[\xpzc_{t}|\ypzc_{t}] \defeq \sum_{\zpzc_{t}}\ProbP_{F}[\xpzc_{t}\|\zpzc_{t}]\ProbP_{S}[\zpzc_{t}\|\ypzc_{t}]$.
Here, $\|$ is the Kramer's causal conditioning, which indicate a causal relation between the conditioning and the conditioned histories\cite{Kramer:1998ufa,Permuter:2011jra}.
Using the path representation of the fitnesses, we can define the time-backward retrospective path probabilities as
\begin{align}
\ProbP^{i}_{B}[\xpzc_{t},\zpzc_{t}|\ypzc_{t}] &\defeq e^{\LF[\xpzc_{t},\ypzc_{t}]-\CMF^{i}[\ypzc_{t}]}\ProbP_{F}[\xpzc_{t}\|\zpzc_{t}]\ProbP_{S}[\zpzc_{t}\|\ypzc_{t}] \label{eq:PBi},\\
\ProbP^{c}_{B}[\xpzc_{t}|\ypzc_{t},\zpzc_{t}] &\defeq e^{\LF[\xpzc_{t},\ypzc_{t}]-\CMF^{c}[\ypzc_{t},\zpzc_{t}]}\ProbP_{F}[\xpzc_{t}\|\zpzc_{t}], \label{eq:PBc},
\end{align}
where $\ProbP^{i}_{B}$ and $\ProbP^{c}_{B}$ are the probabilities of observing a phenotypic history $\xpzc_{t}$ when we trace the phenotypic history in a time-backward manner, retrospectively\cite{Sughiyama:2015cf,Kobayashi:2015eca,Kobayashi:2017tk}. 
In contrast, $\ProbP_{F}[\xpzc_{t}\|\zpzc_{t}]$ is the probability of observing $\xpzc_{t}$ when we trace the phenotypic history in a time forward manner\cite{Sughiyama:2015cf,Kobayashi:2015eca,Kobayashi:2017tk}. 
The difference between the two is attributed to the impact of selection, which can be characterized by investigating a population after selection, retrospectively.

\section{Stochastic trajectories of individual and common sensing}
\begin{figure}
\includegraphics[width=\linewidth]{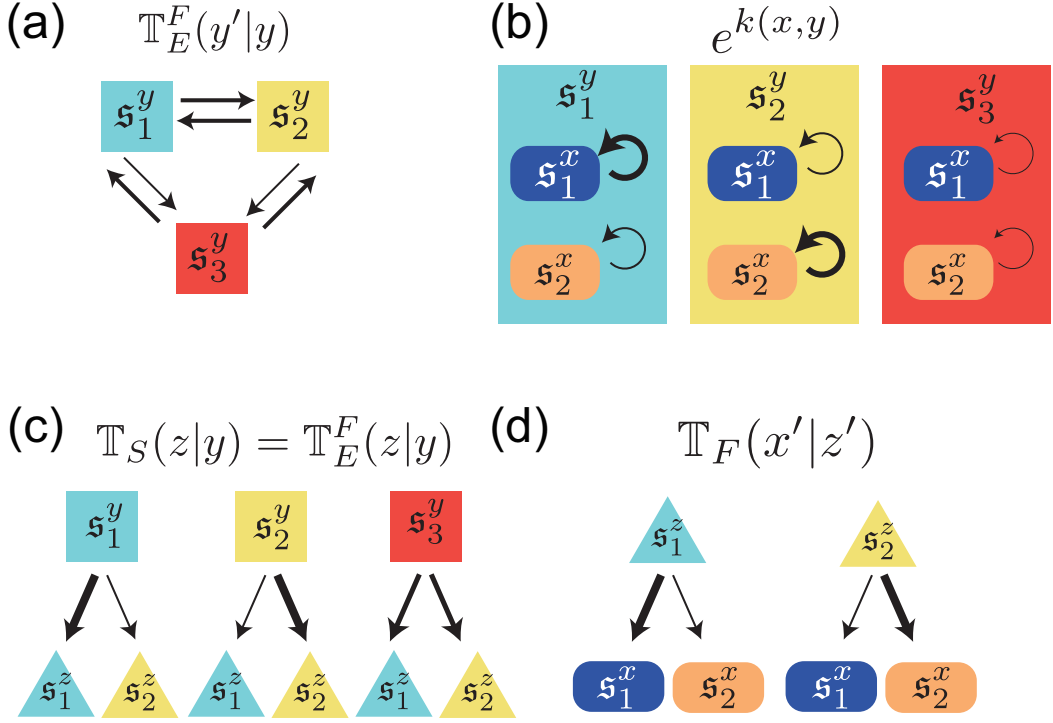}
\caption{\label{fig2} (a) A diagrammatic representation of state transitions of the environment. (b) Replication rates of cells with different phenotypic states under different environmental states. (c) Environment-dependence of the sensing signal. (d) Signal-dependent phenotype switching. The thickness of arrows represent relative probabilities and rates of replications. 
The values of the parameters used for the simulation are given by Eqns. (\ref{eq:T1})--(\ref{eq:T4}).
}
\end{figure}
In order to provide numerical examples of the difference between individual and common sensing, we consider a Markovian environment with three states, $\sety=\{\staty_{1}, \staty_{2}, \staty_{3}\}$, and a population with two phenotypic states, $\setx=\{\statx_{1}, \statx_{2}\}$.
Of the three environmental states, $\staty_{1}$ and $\staty_{2}$ are nutrient A- and nutrient B-rich environments, respectively.
The environmental states fluctuate between these two states, most of time (\com{Fig. 2 (a)}).
In contrast, $\staty_{3}$ is a nutrient-poor environment, in which the growth of the population is limited (\com{Fig. 2 (b)}).
The environmental state occasionally sojourns in this state from either $\staty_{1}$ or $\staty_{2}$ (\com{Fig. 2 (a)}).
The rule for these stochastic transitions among the environmental states is specified by a stochastic transition matrix, $\ProbT_{E}^{F}(y'|y)$, from $y$ to $y'$:
\begin{align}
\{\ProbT_{E}^{F}(y'|y)\}=\bordermatrix{     & \staty_{1} & \staty_{2} & \staty_{3} \cr
               \staty_{1} & 0.70 & 0.25 & 0.25 \cr
               \staty_{2} & 0.25 & 0.70 & 0.25  \cr
               \staty_{3} & 0.05 & 0.05 & 0.50  \cr
            }.\label{eq:T1}
\end{align}
The two phenotypic states, $\statx_{1}$ and $\statx_{2}$, are assumed to be adapted specifically to the nutrient A-rich state $\staty_{1}$ and the nutrient B-rich state $\staty_{2}$, respectively. These are modeled by the replication rates $\lf(\statx_{1},\staty_{1})$ and $\lf(\statx_{2},\staty_{2})$ in the adaptive environments, which are higher than those of $\lf(\statx_{1},\staty_{2})$ and $\lf(\statx_{2},\staty_{1})$ in the non-adaptive environment (\com{Fig. 2 (b)}):
\begin{align}
\{e^{\lf(x,y)}\}=\bordermatrix{     & \staty_{1} & \staty_{2} & \staty_{3} \cr
               \statx_{1} & 2.24 & 0.32 & 0.08 \cr
               \statx_{2} & 0.32 & 2.24 & 0.08  \cr
            }.\label{eq:T2}
\end{align}
The sensing signal has two states, $\setz=\{\statz_{1}, \statz_{2}\}$, which correspond to the nutrient A- and nutrient B-rich environments, $\staty_{1}$ and $\staty_{2}$, respectively.
A cell in the case of individual sensing, or cells in the case of the common sensing, receive $\statz_{1}$ and $\statz_{2}$  with high probability when the environmental state is $\staty_{1}$ or $\staty_{2}$, respectively.
If the environment is in the nutrient-poor $\staty_{3}$ state, a cell or cells obtain $\statz_{1}$ or $\statz_{2}$ with equal probability.
Here, the sensing is assumed to be memory-less, and, thus, its stochastic behavior is defined by a transition matrix, $\ProbT_{S}(z|y)$, for  individual sensing, and by $\ProbT_{E}^{F}(z|y)$ for common sensing (\com{Fig. 2 (c)}):
\begin{align}
\{\ProbT_{S}(z|y)\}=\{\ProbT_{E}^{F}(z|y)\}=\bordermatrix{     & \staty_{1} & \staty_{2} & \staty_{3} \cr
               \statz_{1} & 0.8 & 0.2 & 0.5 \cr
               \statz_{2} & 0.2 & 0.8 & 0.5  \cr
            }.\label{eq:T3}
\end{align}
In order to compare individual and common sensing, we set the accuracy of sensing to be equal, $\ProbT_{S}(z|y) = \ProbT_{E}^{F}(z|y)$,  for all $y\in \sety$ and $z \in \setz$.
Finally, a cell is assumed to switch into phenotypic state $\statx_{i}$ with high probability when it receives sensing signal $\statz_{i}$ for $i=\{1,2\}$ (\com{Fig. 2 (d)}):
\begin{align}
\{\ProbT_{F}(x'|z)\}=\bordermatrix{     & \statz_{1} & \statz_{2} \cr
               \statx_{1} & 0.95 & 0.05  \cr
               \statx_{2} & 0.05 & 0.95   \cr
            },\label{eq:T4}
\end{align}
where the phenotypic switching is set to be memory-less $\ProbT_{F}(x'|x,z)=\ProbT_{F}(x'|z)$.

\begin{figure}
\includegraphics[width=\linewidth]{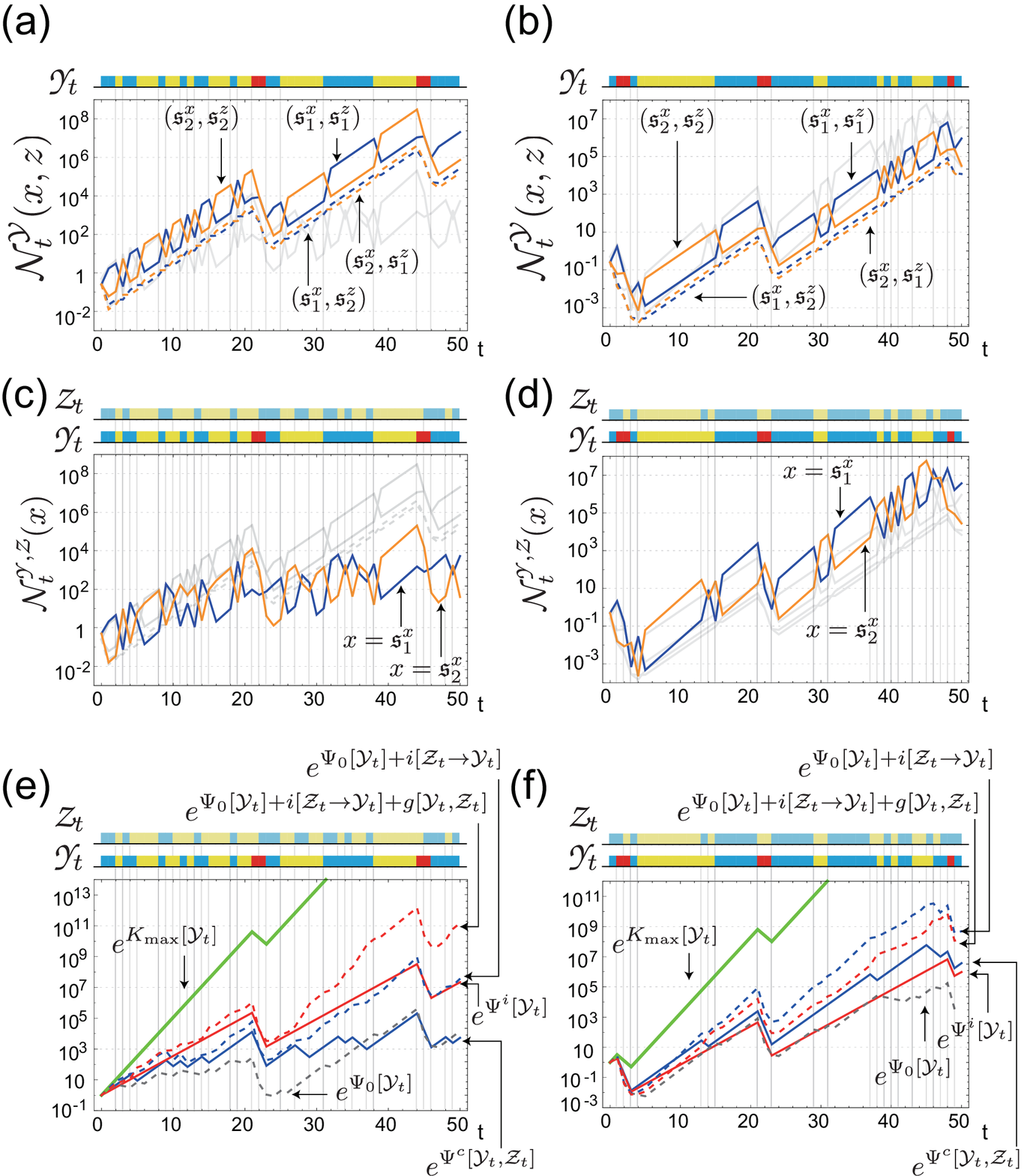}
\caption{\label{fig3} (a,b) Trajectories of populations with individual sensing under two different realizations of the environment. 
Each line corresponds to the population size of the cells with phenotypic state $x$ and sensing signal $z$; the actual value of $(x, z)$ is designated in the panels.
(c,d) Trajectories of populations with common sensing under the same realizations of the environment as in (a) and (b), respectively.
Each line corresponds to the population size of the cells with phenotypic state $x$, with the actual value of $x$ designated in the panels.
(e,f) Fitnesses of the populations with the individual and the common sensing,  $\CMF^{i}[\ypzc_{t}]$ (red solid curve) and $\CMF^{c}[\ypzc_{t},\zpzc_{t}]$ (blue solid curve) under the same realizations of the environment and common signal as in (a,c) and (b,d). Related quantities are also shown for comparison.
}
\end{figure}

Given these conditions, \com{Figure 3} illustrates  the population dynamics of cells with individual sensing (a,b) and with common sensing (c,d) under two different realizations of the environment.
For the first realization, shown in \com{Fig. 3 (a,c,e)}, $\CMF^{i}[\ypzc_{t}]$ is higher than $\CMF^{c}[\ypzc_{t},\zpzc_{t}]$ (see red and blue solid lines in \com{Fig.3 (e)}), whereas, for the second realization (\com{Fig. 3 (b,d,f)}),  $\CMF^{c}[\ypzc_{t},\zpzc_{t}]$ is greater than $\CMF^{i}[\ypzc_{t}]$ (\com{Fig.3 (f)}).
This clearly illustrates that the fitness advantages of individual and common sensing are strongly dependent on the actual realization of the environment and the common sensing signal.
When common sensing produces a correct signal by chance, the population with common sensing can enjoy a higher fitness gain than that with individual sensing.
However, the population with common sensing loses fitness when the signal is incorrect.
\com{Figure 4} also shows the behaviors of $\CMF^{i}[\ypzc_{t}]$ (\com{Fig. 4 (b)}) and $\CMF^{c}[\ypzc_{t},\zpzc_{t}]$ (\com{Fig. 4 (c)}) under $100$ different realizations of $\{\ypzc_{t},\zpzc_{t}\}$, which reinforces the observation that  both $\CMF^{i}[\ypzc_{t}]$ and $\CMF^{c}[\ypzc_{t},\zpzc_{t}]$ can fluctuate significantly, depending on the realizations.
However, an ensemble average of the fitness show that $\average{\Psi^{i}}_{\ProbQ}$ is greater than $\average{\Psi^{c}}_{\ProbQ}$, at least for this specific instance (the red and blue solid lines in\com{Fig. 4 (a)}).

\begin{figure}
\includegraphics[width=\linewidth]{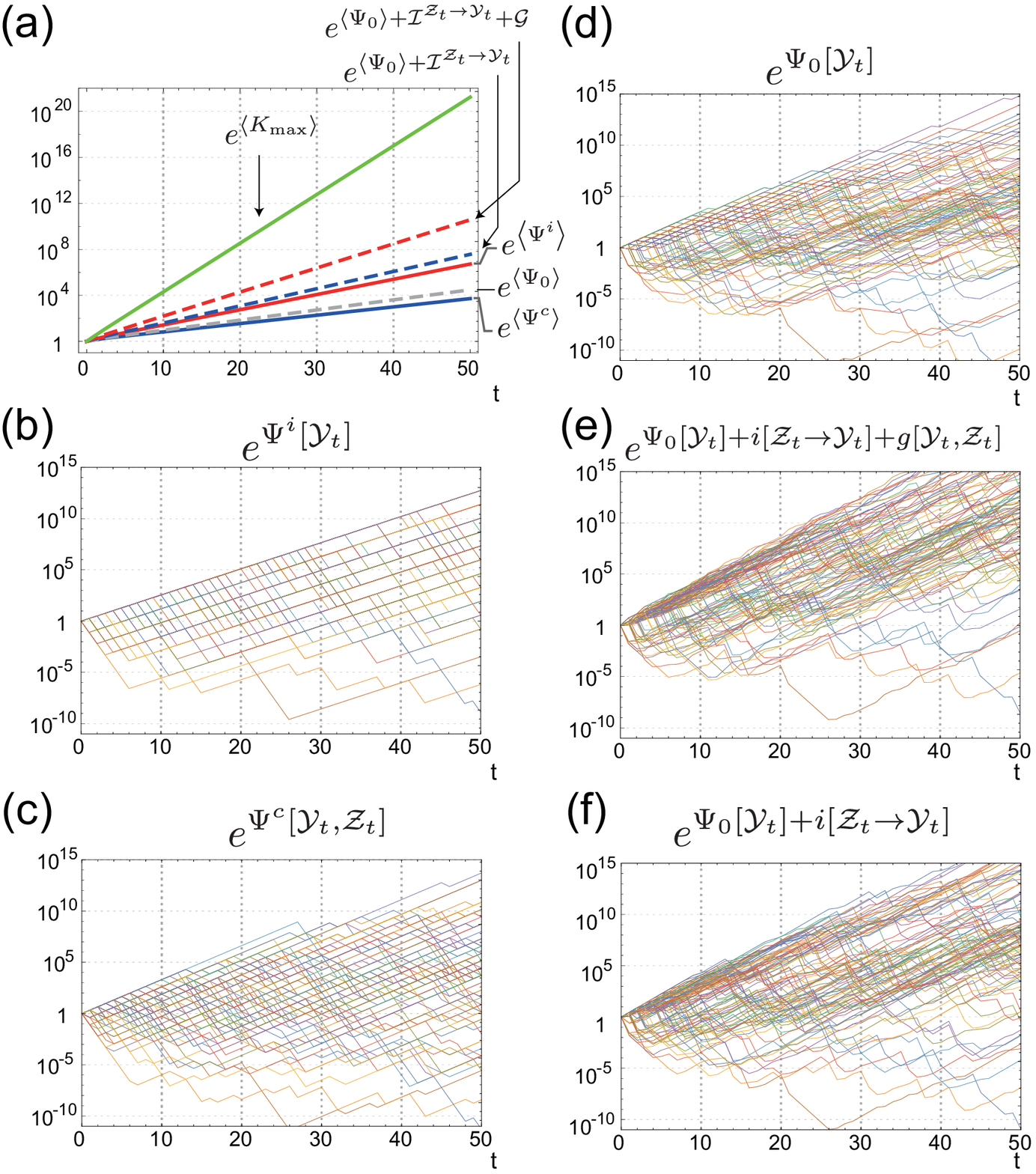}
\caption{\label{fig4} (a) Average values of fitnesses and related quantities. (b, c) Fluctuation of the fitness with individual sensing $\CMF^{i}[\ypzc_{t}]$ (b) and  that with common sensing $\CMF^{c}[\ypzc_{t}]$ (c). (d,e,f) Fluctuation of $\CMF_{0}[\ypzc_{t}]$ (d), $\CMF_{0}[\ypzc_{t}] + i[\zpzc_{t} \to \ypzc_{t}] + \isgain[\ypzc_{t},\zpzc_{t}]$ (e), and $\CMF_{0}[\ypzc_{t}] + i[\zpzc_{t} \to \ypzc_{t}]$ (f).}
\end{figure}


\section{Value of individual sensing is ALWAYS greater than that of common sensing}
In order to characterize the fitness difference between individual and common sensing in general, $\isgain[\ypzc_{t},\zpzc_{t}]\defeq \CMF^{i}[\ypzc_{t}]-\CMF^{c}[\ypzc_{t},\zpzc_{t}]$, 
 we derive a detailed fluctuation relation for the fitness difference $\isgain[\ypzc_{t},\zpzc_{t}]$ from Eqs. (\ref{eq:PBi}--\ref{eq:PBc}) as
\begin{align}
e^{-\isgain[\ypzc_{t},\zpzc_{t}]}=\frac{\ProbP^{i}_{B}[\xpzc_{t},\zpzc_{t}|\ypzc_{t}]}{\ProbP^{c}_{B}[\xpzc_{t}|\ypzc_{t},\zpzc_{t}]\ProbP_{S}[\zpzc_{t}\|\ypzc_{t}] }=\frac{\ProbP^{i}_{B}[\zpzc_{t}|\ypzc_{t}]}{\ProbP_{S}[\zpzc_{t}\|\ypzc_{t}]},\label{eq:DFRg}
\end{align}
where $\ProbP^{i}_{B}[\zpzc_{t}|\ypzc_{t}] \defeq \sum_{\xpzc_{t}}\ProbP^{i}_{B}[\xpzc_{t},\zpzc_{t}|\ypzc_{t}]$.
By assuming that the statistical property of common sensing is the same as that of individual sensing, $\ProbQ[\zpzc_{t}\|\ypzc_{t}]=\ProbP_{S}[\zpzc_{t}\|\ypzc_{t}]$, as in \com{Figs. 3 and 4},  
we obtain the average fluctuation relation as
\begin{align*}
\average{\CMF^{i}[\ypzc_{t}]}_{\ProbQ[\ypzc_{t}]}-\average{\CMF^{c}[\ypzc_{t},\zpzc_{t}]}_{\ProbQ[\ypzc_{t},\zpzc_{t}]}=\ISgain,
\end{align*}
where 
\begin{align}
\ISgain \defeq \average{\isgain}_{\ProbQ}=\KL{\ProbP_{S}[\zpzc_{t}\|\ypzc_{t}]\ProbQ[\ypzc_{t}]}{\ProbP^{i}_{B}[\zpzc_{t}|\ypzc_{t}]\ProbQ[\ypzc_{t}]}
\end{align}
 is the Kulback--Leibler (KL) divergence between the time-forward sensing behavior, $\ProbP_{S}[\zpzc_{t}\|\ypzc_{t}]$, and the time-backward behavior, $\ProbP^{i}_{B}[\zpzc_{t}|\ypzc_{t}]$.
Together with the non-negativity of the KL divergence, the average FR indicates that the average fitness of individual sensing is always greater than that of common sensing by $\ISgain\ge0$. 
Because individual and common sensing are assumed to have the same statistical property, the source of the gain $\ISgain$ is attributed to the individuality of the sensing. 
In the case of individual sensing, the organisms receiving the correct signal by chance  grow more than those that receive incorrect signal do.
Thus, the retrospective signal histories $\ProbP^{i}_{B}[\zpzc_{t}|\ypzc_{t}]$ are biased by the selection from the time-forward signal histories $\ProbP_{S}[\zpzc_{t}\|\ypzc_{t}]$. 
The gain $\ISgain$ is exactly this bias, quantified by the KL divergence. 
No such gain is obtained from the common sensing, because the sensing signal is common to all organisms and, thus, no bias is induced by selection.
This result clearly indicates that the fitness value of individual sensing cannot be properly evaluated by considering only the time-forward behavior of the signal and the environment. 
Whereas individual sensing gains more fitness than common sensing does, on average, as demonstrated in \com{Fig. 3}, $g[\ypzc_{t},\zpzc_{t}]$ can fluctuate significantly and common sensing can gain more fitness than individual sensing does, by chance (\com{Fig. 3 (b) and (d)}).
From the detailed FR for $g[\ypzc_{t},\zpzc_{t}]$ (\eqnref{eq:DFRg}), we also derive the integral fluctuation relation:
\begin{align*}
\average{e^{-\isgain[\ypzc_{t},\zpzc_{t}]}}_{\ProbQ[\ypzc_{t},\zpzc_{t}]}=\average{e^{-(\CMF^{i}[\ypzc_{t}]-\CMF^{c}[\ypzc_{t},\zpzc_{t}])}}_{\ProbQ[\ypzc_{t},\zpzc_{t}]}=1,
\end{align*}
which clarifies that $g[\ypzc_{t},\zpzc_{t}]$ fluctuates, such that the positive $g[\ypzc_{t},\zpzc_{t}]$ balances the negative $g[\ypzc_{t},\zpzc_{t}]$ to satisfy the equality. 
The integral FR is also verified numerically in \com{Fig. 5 (a) and (b)}.

\begin{figure}
\includegraphics[width=\linewidth]{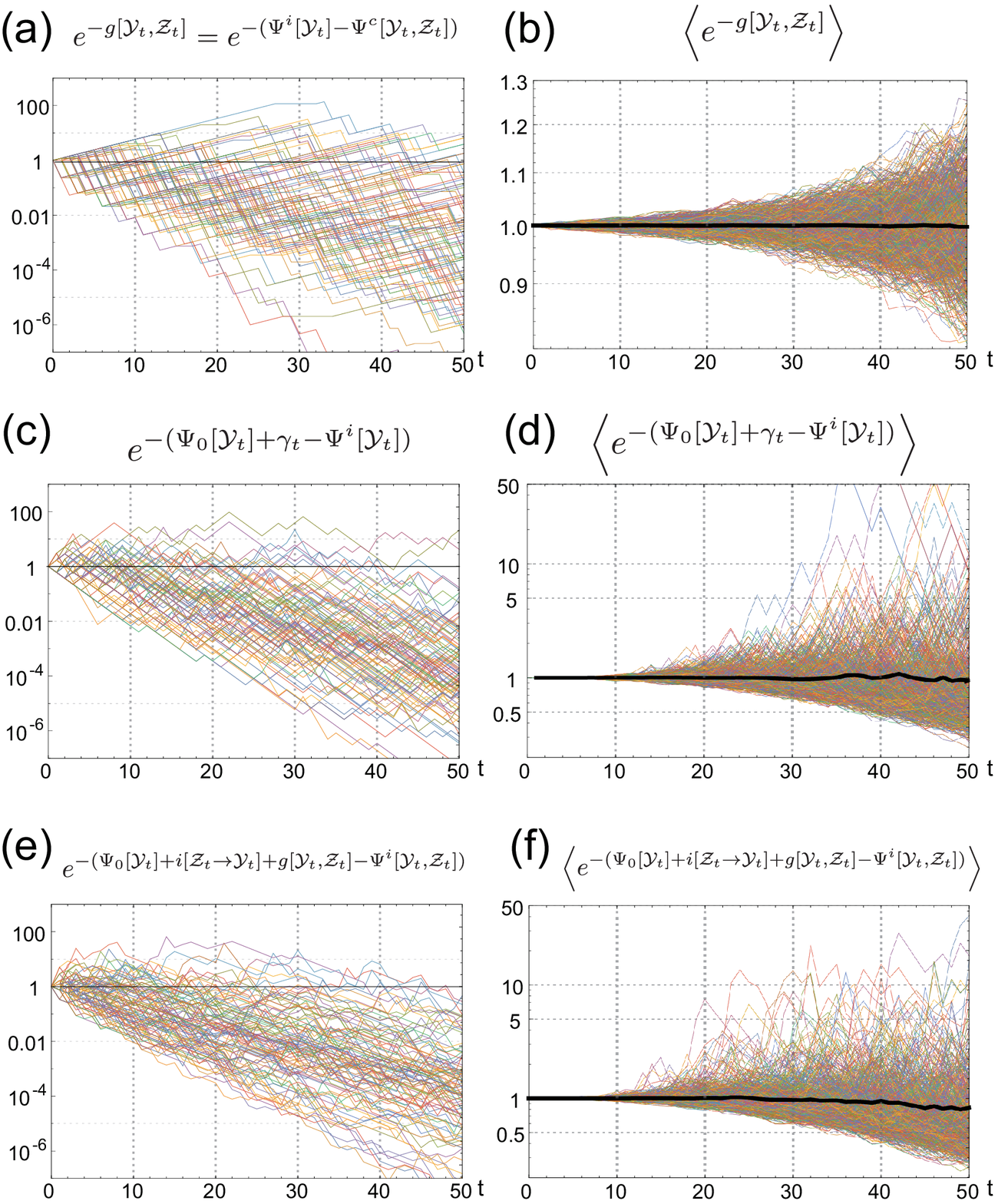}
\caption{\label{fig5} Numerical verification of IFRs for $\isgain[\ypzc_{t}]$ (a and b), $\gamma_{t}-\sigma[\ypzc_{t}]$ (c and d), and $\CMF_{0}[\ypzc_{t}]+i[\zpzc_{t}\to\ypzc_{t}]+g[\ypzc_{t},\zpzc_{t}]-\CMF^{i}[\ypzc_{t}]$ (e and f). 
Left panels are behaviors of the integrands of the IFRs for $100$ different realizations of the environmental and common signal histories.
Right panels are the sample averages of the integrands of the IFRs. 
Thin colored curves are obtained by averaging $10^{5}$ different samples, and the thick black curves are obtained by the average of $1.2 \times 10^{8}$ samples. 
}
\end{figure}

\subsection{The gain of fitness by individual sensing}
We further investigate $\CMF^{i}[\ypzc_{t}]$ to clarify how the fitness of the organisms with individual sensing is shaped.
To this end, as in a previous work \cite{Kobayashi:2017tk}, which investigated the fitness value of common sensing, we additionally assume that $\lf(x,y)$ can be decomposed as $e^{\lf(x,y)}=e^{\lf_{max}(y)}\ProbT_{\LF}(y|x)$\cite{Kobayashi:2017tk}. 
There, $\lf_{max}(y)$ is the maximum replication rate attained if the organisms allocate all their metabolic resources to adapt only to the environmental state $y$.
Therefore, the organisms die out under the environmental states other than $y$. 
$\ProbT_{\LF}(y|x)$ is the fraction of metabolic resources allocated to the environmental state $y$ by a phenotypic state $x$, which defines the metabolic allocation strategy of the organisms.
By defining 
\begin{align}
\ProbP_{\LF}[\ypzc_{t}\|\xpzc_{t}] &\defeq \prod_{\tau=0}^{t-1}\ProbT_{\LF}(y_{\tau+1}|x_{\tau+1}),\\
\LF_{\max}[\ypzc_{t}] &\defeq \sum_{\tau=1}^{t}\lf_{\max}(y_{\tau}),
\end{align}
the historical fitness is decomposed as 
\begin{align}
\LF[\xpzc_{t},\ypzc_{t}] = \LF_{\max}[\ypzc_{t}] + \ln \ProbP_{\LF}[\ypzc_{t}\|\xpzc_{t}].
\end{align} 
By introducing this decomposition into \eqnref{eq:PBi}, we obtain
\begin{align}
e^{\CMF^{i}[\ypzc_{t}]-\CMF_{0}[\ypzc_{t}]}= \frac{\ProbP_{\LF}[\ypzc_{t}\|\xpzc_{t}]\ProbP_{F}[\xpzc_{t}\|\zpzc_{t}]\ProbP_{S}[\zpzc_{t}\|\ypzc_{t}]}{\ProbP^{i}_{B}[\xpzc_{t},\zpzc_{t}|\ypzc_{t}]\ProbQ[\ypzc_{t}]},\label{eq:FitnessDecomp1}
\end{align}
where $\CMF_{0}[\ypzc_{t}]\defeq \LF_{\max}[\ypzc_{t}] + \ln \ProbQ[\ypzc_{t}]$, the average of which is known to bound the average fitness of a population without sensing\cite{Kobayashi:2017tk}.
By taking the marginalization with respect to $\xpzc_{t}$ and $\zpzc_{t}$, we have
\begin{align}
\CMF^{i}[\ypzc_{t}]= \CMF_{0}[\ypzc_{t}] + \sigma[\ypzc_{t}]= \LF_{\max}[\ypzc_{t}] + \ln \ProbP_{\LF F S}[\ypzc_{t}|\ypzc_{t}],\label{eq:PsiI}
\end{align}
where  
\begin{align*}
\ProbP_{\LF F S}[\ypzc'_{t}|\ypzc_{t}] &\defeq \sum_{\xpzc_{t},\zpzc_{t}}\ProbP_{\LF}[\ypzc'_{t}\|\xpzc_{t}]\ProbP_{F}[\xpzc_{t}\|\zpzc_{t}]\ProbP_{S}[\zpzc_{t}\|\ypzc_{t}],
\end{align*}
and
\begin{align*}
\sigma[\ypzc_{t}] &\defeq \ln \frac{\ProbP_{\LF F S}[\ypzc_{t}|\ypzc_{t}]}{\ProbQ[\ypzc_{t}]}.
\end{align*}
Because the average of $\CMF_{0}[\ypzc_{t}]$ is the tight bound of the fitness without sensing, $\sigma[\ypzc_{t}]$ is the gain in fitness from individual sensing.
Here, $\ProbP_{\LF F S}[\ypzc'_{t}|\ypzc_{t}]$ is the probability that an organism allocates its metabolic resources to an environmental history $\ypzc'_{t}$ when it experiences environmental history $\ypzc_{t}$.
Thus, $\ProbP_{\LF F S}[\ypzc_{t}|\ypzc_{t}]$ measures the probability that the metabolic resource is correctly allocated to the actual environmental history $\ypzc_{t}$, and $1-\ProbP_{\LF F S}[\ypzc_{t}|\ypzc_{t}]$ is the probability of an incorrect allocation.
In other wards, $\ProbP_{\LF F S}[\ypzc_{t}|\ypzc_{t}]$ characterizes how accurately the individual sensing, phenotypic switching, and metabolic allocation together respond to the actual environment.
From an information-theoretic viewpoint, this cascade from environment to metabolic allocation via sensing and phenotypic switching is very similar to the auto-encoding and decoding of information $\ypzc_{t}$ via multiple layers \cite{pmlr-v27-baldi12a}. 
The sensing works as the encoding of an environmental history $\ypzc_{t}$ into $\zpzc_{t}$. 
The signal-dependent phenotypic switching is the processing of the encoded signal in the internal layers.
The metabolic allocation is the decoding process to recover the original information, $\ypzc_{t}$, from $\xpzc_{t}$.
Under this interpretation, $\ProbP_{\LF F S}[\ypzc'_{t}|\ypzc_{t}]$ determines the statistical correspondence between the encoded information $\ypzc_{t}$ and the decoded information $\ypzc'_{t}$, and $\ProbP_{\LF F S}[\ypzc_{t}|\ypzc_{t}]$ is the probability that the encoded data $\ypzc_{t}$ is correctly decoded as $\ypzc_{t}$.
Therefore, the total fidelity can be quantified as
\begin{align}
\gamma_{t}\defeq \ln \sum_{\ypzc_{t}}\ProbP_{\LF F S}[\ypzc_{t}|\ypzc_{t}] = \ln\average{e^{\sigma[\ypzc_{t}]}}_{\ProbQ[\ypzc_{t}]}.\label{eq:gamma}
\end{align}
Formally, the same quantities, $\sigma[\ypzc_{t}]$ and $\gamma_{t}$, were introduced by Sagawa and Ueda as the coarse-grained entropy production and the efficiency parameter of feedback control in information thermodynamics\cite{Sagawa:2012wi}.
Using $\gamma_{t}$, $\sigma[\ypzc_{t}]$ can be decomposed as
\begin{align*}
\sigma[\ypzc_{t}]=\gamma_{t} - \ln\frac{\ProbQ[\ypzc_{t}]}{\ProbP_{\gamma}[\ypzc_{t}]},
\end{align*}
where  
\begin{align}
\ProbP_{\gamma}[\ypzc_{t}] \defeq e^{-\gamma_{t}}\ProbP_{\LF F S}[\ypzc_{t}|\ypzc_{t}], \label{eq:Pgamma}
\end{align}
is a path probability.
By combining this with \eqnref{eq:PsiI}, we have
\begin{align}
\CMF^{i}[\ypzc_{t}]= \CMF_{0}[\ypzc_{t}] + \gamma_{t} - \ln\frac{\ProbQ[\ypzc_{t}]}{\ProbP_{\gamma}[\ypzc_{t}]}.\label{eq:DFRsigma}
\end{align}
By taking the average with respect to $\ProbQ[\ypzc_{t}]$, we obtain  
\begin{align}
\average{\CMF^{i}}_{\ProbQ}=\average{\CMF_{0}}_{\ProbQ}+\gamma_{t} - \KL{\ProbQ[\ypzc_{t}]}{\ProbP_{\gamma}[\ypzc_{t}]} \le \average{\CMF_{0}}_{\ProbQ}+\gamma_{t}.\label{eq:AFRsigma}
\end{align}
Equations (\ref{eq:DFRsigma}) and (\ref{eq:AFRsigma}) can be regarded as detailed and average FRs, respectively, with respect to $\CMF_{0}[\ypzc_{t}]+\gamma_{t} - \CMF^{i}[\ypzc_{t}]$.
Because $\average{\CMF_{0}}_{\ProbQ}$ is the tight upper bound of the average fitness without sensing, this relation means that $\gamma_{t}$ is an upper bound of the fitness gain from individual sensing.
Moreover, $\gamma_{t}$ is an intrinsic quantity of the population, in the sense that it is determined irrespective of the actual statistical law of the environment, $\ProbQ[\ypzc_{t}]$.
The deviation of $\CMF^{i}[\ypzc_{t}]$ from $\average{\CMF_{0}}_{\ProbQ}+\gamma_{t}$ satisfies an integral FR as
\begin{align}
\average{e^{-(\CMF_{0}[\ypzc_{t}] + \gamma_{t}-\CMF^{i}[\ypzc_{t}])}}_{\ProbQ[\ypzc_{t}]}=\average{e^{-(\gamma_{t}-\sigma[\ypzc_{t}])}}_{\ProbQ[\ypzc_{t}]}=1,\label{eq:IFRsigma}
\end{align}
the behaviors of which are illustrated numerically in (\com{Fig. 5 (c) and (d)}).

\subsection{Connection with Other Information Measures}
In order to link the quantities $\sigma$ and $\gamma_{t}$ with other common information measures, we further assume that the environment is Markovian: 
\begin{align}
\ProbQ[\ypzc_{t}]=\prod_{\tau=0}^{t-1} \ProbT_{E}^{F}(y_{\tau+1}|y_{\tau})p_{E}(y_{0}), 
\end{align}
and that the sensing is memory less as
\begin{align}
\ProbT_{S}(z_{t+1}|z_{t},y_{t+1})=\ProbT_{S}(z_{t+1}|y_{t+1}).
\end{align}
Then, we obtain the joint time-forward probability for $\ypzc_{t}$ and $\zpzc_{t}$ and its Bayesian causal decomposition as
\begin{align*}
\ProbP_{S}[\ypzc_{t},\zpzc_{t}] &\defeq \ProbP_{S}[\zpzc_{t}\|\ypzc_{t}]\ProbQ[\ypzc_{t}] =\ProbP_{S}^{B}[\ypzc_{t}\|\zpzc_{t}]\ProbP_{S}^{B}[\zpzc_{t}\|\ypzc_{t-1}],
\end{align*}
where
\begin{align}
\ProbP_{S}^{B}[\ypzc_{t}\|\zpzc_{t}] &\defeq \prod_{t=0}^{t-1}\ProbT_{E}^{B}(y_{t+1}|z_{t+1}, y_{t})p(y_{0}|z_{0}),\\
\ProbP_{S}^{B}[\zpzc_{t}\|\ypzc_{t-1}] &\defeq \prod_{t=0}^{t-1}\ProbT_{E}^{B}(z_{t+1}|y_{t})p(z_{0})
\end{align}
 are path probabilities generated by the Bayesian sequential inference, defined as
\begin{align}
\ProbT_{E}^{B}(z_{t+1}|y_{t}) & \defeq \sum_{y_{t+1}}\ProbT_{S}(z_{t+1}|y_{t+1})\ProbT_{E}^{F}(y_{t+1}|y_{t}),\\
\ProbT_{E}^{B}(y_{t+1}|z_{t+1}, y_{t}) &\defeq \frac{\ProbT_{E}^{F}(z_{t+1}|y_{t+1})\ProbT_{E}^{F}(y_{t+1}|y_{t})}{\ProbT_{E}^{B}(z_{t+1}|y_{t})},
\end{align}
where $\ProbT_{E}^{B}(y_{t+1}|z_{t+1}, y_{t})$ is the Bayesian posterior of the environmental state, $y_{t+1}$, given the information of the sensed signal $z_{t+1}$ and the previous environmental state $y_{t}$.
Then, by using \eqnref{eq:DFRg}, \eqnref{eq:FitnessDecomp1} can be rearranged as
\begin{align}
e^{-\left(\CMF^{i}[\ypzc_{t}]-(\CMF_{0}[\ypzc_{t}] + i[\zpzc_{t} \to \ypzc_{t}] + \isgain[\ypzc_{t},\zpzc_{t}])\right)} =\frac{\ProbP_{S}^{B}[\ypzc_{t}\|\zpzc_{t}]}{\ProbP_{\LF,F}[\ypzc_{t}|\zpzc_{t}]}, \label{eq:DFRI}
\end{align}
where $\ProbP_{\LF,F}[\ypzc_{t}|\zpzc_{t}]:=\sum_{\xpzc_{t}}\ProbP_{\LF}[\ypzc_{t}\|\xpzc_{t}]\ProbP_{F}[\xpzc_{t}\|\zpzc_{t}]$ and $i[\zpzc_{t} \to \ypzc_{t}]  \defeq \ln \ProbP_{S}^{B}[\ypzc_{t}\|\zpzc_{t}]/\ProbQ[\ypzc_{t}]$ is the pointwise directed information from $\zpzc_{t}$ to $\ypzc_{t}$.
This is another detailed FR with individual sensing, the average version of which can be obtained by taking the average with respect to $\ProbP_{S}[\ypzc_{t}, \zpzc_{t}]$:
\begin{align}
\average{\CMF^{i}}_{\ProbQ} =\average{\CMF_{0}}_{\ProbQ} + \MI^{\zpzc_{t} \to \ypzc_{t}} + \ISgain- \KLD_{\loss}, \label{eq:indPsiav}
\end{align}
where $\KLD_{\loss}=\KL{\ProbP_{S}[\ypzc_{t},\zpzc_{t}]}{\ProbP_{\LF,F}[\ypzc_{t}|\zpzc_{t}]\ProbP_{S}^{B}[\zpzc_{t}\|\ypzc_{t-1}]}$ and $\MI^{\zpzc_{t} \to \ypzc_{t}} \defeq \average{i[\zpzc_{t} \to \ypzc_{t}] }_{\ProbP_{S}[\ypzc_{t},\zpzc_{t}]}$ is the directed information\cite{Permuter:2011jra}.
Their integral version is illustrated numerically in \com{Fig. 5 (e) and (f)}.
Because $\isgain[\ypzc_{t},\zpzc_{t}]=\CMF^{i}[\ypzc_{t}]-\CMF^{c}[\ypzc_{t},\zpzc_{t}]$, 
we can immediately see that Eqns (\ref{eq:DFRI}) and (\ref{eq:indPsiav}) are exactly equivalent to the detailed and average FRs, respectively, for the fitness with common sensing:
\begin{align}
e^{-\left(\CMF^{c}[\ypzc_{t}, \zpzc_{t}]-(\CMF_{0}[\ypzc_{t}] + i[\zpzc_{t} \to \ypzc_{t}])\right)} =\frac{\ProbP_{S}^{B}[\ypzc_{t}\|\zpzc_{t}]}{\ProbP_{\LF,F}[\ypzc_{t}|\zpzc_{t}]}, \label{eq:DFRc}
\end{align}
and
\begin{align}
\average{\CMF^{c}}_{\ProbQ} =\average{\CMF_{0}}_{\ProbQ} + \MI^{\zpzc_{t} \to \ypzc_{t}} - \KLD_{\loss}. \label{eq:AFRc}
\end{align}
These relations were originally derived in ref\cite{Kobayashi:2017tk}.
For a given and fixed sensing property, $\ProbT_{S}(z_{\tau}|y_{\tau})$, 
the maximum gain of the average fitness by common sensing is shown to be bounded by $\MI^{\zpzc_{t} \to \ypzc_{t}}$ as
\begin{align}
\max_{\ProbT_{F},\ProbT_{\LF}}\average{\CMF^{c}}_{\ProbQ} -  \average{\CMF_{0}}_{\ProbQ} \le \MI^{\zpzc_{t} \to \ypzc_{t}},
\end{align}
where the equality is attained when $\KLD_{\loss}=0$.
$\KLD_{\loss}$ is the loss of fitness due to an imperfect implementation of a sequential Bayesian inference, and becomes $0$ if and only if the phenotypic switching strategy, $\ProbP_{F}^{*}[\xpzc_{t}\|\zpzc_{t}]$, and the metabolic allocation strategy, $\ProbP_{\LF}^{*}[\ypzc_{t}\|\xpzc_{t}]$, are jointly optimized to implement the Bayesian sequential inference as $\ProbP_{\LF,F}^{*}[\ypzc_{t}|\zpzc_{t}] = \ProbP_{S}^{B}[\ypzc_{t}\|\zpzc_{t}]$, where
\begin{align*}
\ProbP_{\LF,F}^{*}[\ypzc_{t}|\zpzc_{t}] &\defeq \sum_{\xpzc_{t}}\ProbP_{\LF}^{*}[\ypzc_{t}\|\xpzc_{t}]\ProbP_{F}^{*}[\xpzc_{t}\|\zpzc_{t}].
\end{align*}
An instance of the optimal metabolic allocation and phenotypic switching strategies is $\ProbT_{\LF}^{*}(y|x)=\delta_{x,y}$ and $\ProbT_{F}^{*}(x'|x,z)=\left.\ProbT_{E}^{B}(y'|z, y)\right|_{y'=x',y=x}$, when $\setx=\sety$.

In contrast, in the case of individual sensing, the Bayesian inference is no longer optimal, because $\ISgain$ is dependent on the strategies of phenotypic switching and metabolic allocation, and $\{\ProbP_{F}^{*}, \ProbP_{\LF}^{*}\}$ may not be the maximizer of $\ISgain$.
This fact is more clearly shown as
\begin{align}
\max_{\ProbT_{F},\ProbT_{\LF}}\average{\CMF^{i}}_{\ProbQ} \ge \average{\CMF^{i*}}_{\ProbQ}=\average{\CMF_{0}}_{\ProbQ} +\MI^{\zpzc_{t} \to \ypzc_{t}} + \ISgain^{*},
\end{align}
where $\CMF^{i*}$ and $\ISgain^{*}$ are obtained by inserting $\ProbP_{F}^{*}$ and $\ProbP_{\LF}^{*}$ that satisfy $\KLD_{\loss}=0$.
Equivalently, from $\sigma[\ypzc_{t}]=\CMF^{i}[\ypzc_{t}]-\CMF_{0}[\ypzc_{t}]$, we have
\begin{align}
\average{\sigma[\ypzc_{t}]}_{\ProbQ[\ypzc_{t}]} = \MI^{\zpzc_{t} \to \ypzc_{t}} + \ISgain- \KLD_{\loss},
\end{align}
and
\begin{align}
\max_{\ProbT_{F},\ProbT_{\LF}}\average{\sigma[\ypzc_{t}]}_{\ProbQ[\ypzc_{t}]} \ge \MI^{\zpzc_{t} \to \ypzc_{t}} + \ISgain^{*}.
\end{align}
This inequality further indicates that the maximum average fitness gain from individual sensing for a fixed sensing strategy is greater than the directed information plus $\ISgain^{*}$, which means that the sequential Bayesian inference is no longer optimal.
It is optimal in the case of the common sensing because the sensing signal is common and the subsequent phenotypic diversification by following the sequential Bayesian inference can hedge the risk of the error optimally.
In the individual sensing, in contrast, stochastic individual sensing automatically induces a diversification in a population, which makes subsequent diversification by following Bayesian posterior suboptimal and redundant. 
Moreover, the information measure of the sensing, such as directed information, may not be an appropriate quantity to capture the efficiency of the overall decision-making process with individual sensing.

\section{Discussion and Future Works}

These results indicate that $\sigma[\ypzc_{t}]$ and $\gamma_{t}$ are more relevant quantities for characterizing the fitness gain from the individual sensing.
From the average FR of $\sigma[\ypzc_{t}]$:
\begin{align*}
\average{\sigma}_{\ProbQ}=\gamma_{t} - \KL{\ProbQ[\ypzc_{t}]}{\ProbP_{\gamma}[\ypzc_{t}]},
\end{align*}
the maximization of $\average{\sigma}_{\ProbQ}$ is reduced to balancing the maximization of the total fidelity $\gamma_{t}$ and the minimization of $\KL{\ProbQ[\ypzc_{t}]}{\ProbP_{\gamma}[\ypzc_{t}]}$.
Because both $\gamma_{t}$ and $\ProbP_{\gamma}[\ypzc_{t}]$ depend on the actual strategies of organisms, there exists tradeoff between them, in general.

\com{In the analogy of autoencoding and decoding, $\gamma_{t}$ becomes higher when each input $\ypzc_{t}$ is decoded more correctly. 
In contrast, $\KL{\ProbQ[\ypzc_{t}]}{\ProbP_{\gamma}[\ypzc_{t}]}$ is minimized when the relative fidelity for $\ypzc_{t}$ matches the probability, $\ProbQ[\ypzc_{t}]$, that the environmental history $\ypzc_{t}$ appears, because $\ProbP_{\gamma}[\ypzc_{t}]$ measures the relative fidelity of decoding $\ypzc_{t}$, given $\ypzc_{t}$ as encoding information. 
From the definition of $\ProbP_{\gamma}[\ypzc_{t}]$ (\eqnref{eq:Pgamma}), $\ProbP_{\gamma}[\ypzc_{t}]\le e^{-\gamma_{t}}$ must hold for each $\ypzc_{t}$.
If the total fidelity $\gamma_{t}$ is fixed and small enough to satisfy $\max_{\ypzc_{t}}\ProbQ[\ypzc_{t}] \le e^{-\gamma_{t}}$, balancing sensing, phenotypic switching, and metabolic allocation to satisfy $\ProbP_{\gamma}[\ypzc_{t}]=\ProbQ[\ypzc_{t}]$ becomes the optimal strategy to maximize $\average{\sigma}$.
This observation suggests that, under biologically realistic situations with moderate total fidelity,  $\ProbP_{\gamma}[\ypzc_{t}]=\ProbQ[\ypzc_{t}]$ can be regarded as a proxy of the optimal strategy with individual sensing.
If the total fidelity is too large to violate $\max_{\ypzc_{t}}\ProbQ[\ypzc_{t}]<e^{-\gamma_{t}}$, however,
$ \KL{\ProbQ[\ypzc_{t}]}{\ProbP_{\gamma}[\ypzc_{t}]}=0$ cannot be achieved, and more complicated optimization is required.}

These investigations in conjunction with the analogy of the problem with autoencoding and decoding, show that in order to understand the decision-making of cells and organisms with individual sensing, we should consider a joint optimization of sensing, phenotypic switching, and metabolic allocation, rather than an optimization of a part of them with the other fixed and given.
In the evolution of cellular and organismal decision-making, these three factors are concurrently subject to natural selection, and we have to frame this problem appropriately.
This challenge may lead to a deeper understanding of thermodynamics with feedback, because similar quantities to $\sigma[\ypzc_{t}]$ and $\gamma_{t}$ have appeared already in the problem of feedback efficiency in information thermodynamics.
Moreover, the analogy of the problem with auto-encoding may pave the way to link the field of machine learning and deep learning with that of evolutionary biology and optimization.

\begin{acknowledgments}
We acknowledge Yuichi Wakamoto, Takahiro Sagawa, and Takashi Nozoe for their useful discussions. 
This research is supported partially by JST PRESTO Grant Number JPMJPR15E4, Japan, and the 2016 Inamori Research Grants Program, Japan.
\end{acknowledgments}



\section*{References}


%

\end{document}